\documentclass[lettersize,11pt]{article} 
\usepackage{verbatim} 
\usepackage{float}
\usepackage{amsthm} 
\usepackage{amsmath}
\usepackage{amsfonts}
\usepackage{graphicx}
\usepackage{setspace}
\usepackage{yfonts}
\usepackage{paralist} 
\usepackage{xcolor,colortbl}

\newcommand{\mysection}[1]{\vspace*{-4.5mm}\section{#1}\vspace*{-3mm}}

\newcommand{\ignore}[1]{}

\floatstyle{ruled}
\newfloat{algorithm}{h}{lop}
\floatname{algorithm}{}
\restylefloat{algorithm}

\newcommand{\tab}{\hspace*{.5em}}
\newcommand{\twotab}{\hspace*{1.5em}}
\newcommand{\threetab}{\hspace*{2.5em}} 
\newcommand{\fourtab}{\hspace*{3.5em}}

\newtheoremstyle{slplain}
  {0mm}
  {0mm}
  {\itshape}
  {}
  {\bfseries}
  {.}
  { }
  {}

\theoremstyle{slplain}
\newtheorem{theorem}{Theorem} 
     
\newtheorem{lemma}[theorem]{Lemma}         
  
\newtheorem{invariant}[theorem]{Invariant}
\newtheorem{observation}[theorem]{Observation}

\newcommand{\gothic}[1]{\textswab{#1}}
\newcommand{\gL}{\gothic{L}}
\newcommand{\gS}{\gothic{S}}
\newcommand{\gc}{\gothic{c}}

\newcommand{\gp}{\gothic{p}}
\newcommand{\gq}{\gothic{q}}
\newcommand{\gr}{\gothic{r}}

\title{Non-Blocking Doubly-Linked Lists \\ with Good Amortized Complexity}

\author{Niloufar Shafiei\\ 
Department of Electrical Engineering and Computer Science\\ 
York University\\ 
4700 Keele Street,\\
Toronto, Ontario,\\
Canada M3J 1P3\\
niloo@cse.yorku.ca}
\date{}

\begin{document}



\maketitle

\begin{abstract}
We present a new non-blocking doubly-linked list implementation for an asynchronous shared-memory system. 
It is the first such implementation for which an upper bound on amortized time complexity has been proved.
In our implementation, operations access the list via {\it cursors}.
Each cursor is associated with an item in the list and is local to a process.
The implementation supports two update operations, insertBefore and delete, and two move operations, moveRight and moveLeft.
An insertBefore($c$, $x$) operation inserts an item $x$ into the list immediately before the cursor $c$'s location.
A delete($c$) operation removes the item at the cursor $c$'s location
and sets the cursor to the next item in the list. 
The move operations move the cursor one position to the right or left.  
The update operations use single-word Compare\&Swap instructions.
The move operations only read shared memory and never change the state of the data structure.
If all update operations modify different parts of the list, they run completely concurrently.
Let $\dot c(op)$ be the maximum number of active cursors at any one time during the operation $op$.
The amortized complexity of each update operation $op$ is $O(\dot c(op))$ and each move operation is $O(1)$.
We have written a detailed correctness proof and amortized analysis of our implementation.
\end{abstract}



\section{Introduction}
\vspace*{-2mm}

To take advantage of multicore systems, data structures that can be accessed concurrently are essential.
The linked list is one of the most fundamental data structures and has many applications in distributed systems including processor scheduling, memory management and sparse matrix computations 
\cite{scheduling-app, matrix-app, mem-app}.
It is also used as a building block for more complicated data structures such as deques, skip lists and Fibonacci heaps.
In some applications, the list must keep items in sorted order.

We design a concurrent doubly-linked list for asynchronous shared-memory systems
that is {\it non-blocking} (also sometimes called {\it lock-free}): it guarantees some operation will complete in a finite number of steps. 
The first non-blocking {\it singly}-linked list \cite{SLL} was proposed almost two decades ago.
Designing a non-blocking {\it doubly}-linked list was an open problem for a long time.
Doubly-linked lists were implemented using multi-word synchronization primitives that are not widely available \cite{DLL-color, DLL-DCAS}. 
Sundell and Tsigas \cite{DLL-CAS} gave the first implementation from single-word compare\&swap (CAS).
However, 
they give only a sketch of a correctness proof.  
We compare our implementation to theirs in Section~\ref{rw-sec}.

A process accesses our list via a {\it cursor}, which is an object
in the process's local memory that is located at an item in the list. 
Update operations can insert or delete an item at the cursor's location,
and moveLeft and moveRight operations move the cursor to the adjacent item in either direction.
In \cite{DLL-CAS}, move operations sometimes have to perform CAS steps to help updates complete.
In our implementation, move operations only read shared memory, even when there is contention,
 so they do not interfere with one another.
This is a desirable property since moves are more common than updates in many applications.
If all concurrent updates are on disjoint parts of the list, they do not interfere with one another.
Our implementation is  modular and  can be adapted for other updates, such as replacing
 one item by another.
For simplicity, we assume the existence of a garbage collector (such as the one provided in Java) that deallocates objects that are no longer reachable.

In Section \ref{ss-sec}, we give a novel  specification that describes how updates affect cursors and
how a process gets feedback about other processes' updates at the location of its cursor.
We believe this interface makes the list easy to use as a black box.
In our implementation, a cursor becomes {\it invalid} if another process performs an update at its location.
If an operation is called with an invalid cursor, it returns invalidCursor and makes the cursor valid again.
This avoids having a process perform an operation on the wrong item. 
If another process inserts an item before the cursor, it becomes invalid for insertions
only, to ensure
that an item can be inserted between two specific items. 
This makes it easy to maintain a sorted list.
For example, if two processes try to insert 5 and 7 at the same location simultaneously, one fails and  returns invalidCursor.
This avoids inserting 7 and then  5 out of~order.

A concurrent implementation of a data structure is {\it linearizable} \cite{lin} if each operation appears to take place atomically at some time during the operation.
A detailed proof that our implementation is linearizable appears in \cite{tech-rep}.
One of the main challenges is to ensure
the two pointer changes required by an update appear to occur atomically.
Our implementation uses two CAS steps to change the pointers.
Between the two CAS steps, the data structure is temporarily inconsistent.
We design a mechanism for detecting such inconsistencies and concurrent operations behave as if the second change has already occurred.
Using this mechanism, move operations are performed without altering the shared memory.

We give an amortized analysis of our implementation \cite{tech-rep} (excluding garbage collection). 
This is the first amortized analysis for a non-blocking doubly-linked list.
Some parts of our analysis are similar to the amortized analysis of non-blocking trees in \cite{AA-BST}, 
which used a combination of an aggregate analysis and the accounting method.
Here, we simplified the argument using the potential method.
Let $\dot c(op)$ be the maximum number of active cursors at any one time during the operation $op$.
The amortized complexity of each operation $op$ is $O(\dot c(op))$ for updates and $O(1)$ for moves.
To summarize:
\begin{compactitem}
\item We present a non-blocking linearizable doubly-linked list using single-word CAS.
\item Cursors are updated and moved by only reading the shared memory. 
\item The cursors provided by our implementation are robust: 
they can be used to traverse and update the list, even as concurrent operations modify the list.
\item Our implementation and proof are modular and can be adapted for other data structures.
\item Our implementation can easily maintain a sorted list.
\item In our algorithms, the amortized complexity of each update $op$ is $O(\dot c(op))$ and each move is~$O(1)$.
\end{compactitem}

\mysection{Related Work} \label{rw-sec}
In this paper, we focus on non-blocking algorithms, which do not use locks.
There are two general techniques for obtaining non-blocking data structures:
universal constructions (see 
\cite{WFConstruction} for a survey) and transactional memory (see \cite{STM-survey} for a survey). 
Such general techniques are usually less efficient than implementations designed for specific data structures.
Turek, Shasha and Prakash \cite{coOp} and Barnes \cite{lock-free} introduced a technique 
in which processes cooperate to complete operations to ensure non-blocking progress.
Each update operation creates a descriptor object that contains information 
that other processes can use to help complete the update.
This technique has been used for various data structures.
Here, we extend the scheme used in \cite{Brown1, BST} to coordinate processes for tree structures
and the scheme used in \cite{PT} for updates that make more than one change to a Patricia trie.

Doubly-linked lists can also be implemented using $k$-CAS primitives (which modify $k$ locations atomically).
Although $k$-CAS is usually not available in hardware,
there are $k$-CAS implementations from single-word CAS \cite{Harris-MCAS, Luchangco-MCAS, Sundell-MCAS}.
It is not so straightforward  to build a doubly-linked list using $k$-CAS.
Suppose each item is represented by a node with $nxt$ and $prv$ fields that point to the adjacent nodes. 
Suppose a list has four consecutive nodes, $A$, $B$, $C$ and $D$. 
A deletion of $C$ must change $B.nxt$  from $C$ to $D$ and $D.prv$  from $C$ to $B$.
It is {\it not} sufficient for the deletion update these two pointers with a 2-CAS.
If two concurrent deletions remove $B$ and~$C$ in this way, 
$C$ would still be accessible through $A$ after the two deletions.
This problem can be avoided by using 4-CAS to simultaneously update the two pointers and check whether the two pointers of $C$ still point to $B$ and $D$.
Then, the 4-CAS of one of the two concurrent deletions would fail.
The 4-CAS works for updating pointers, but it is not obvious how to detect  invalidation of cursors and update their locations.
For this, the multiword CAS may have to operate on even more words.
The most efficient $k$-CAS implementation \cite{Sundell-MCAS} uses $2k+1$ CAS steps to change $k$ words when there is no contention.
Thus, at least $9$ CAS steps are required for 4-CAS.
Our implementation uses only $5$ CAS steps for contention-free updates.

Valois \cite{SLL} presented the first non-blocking implementation of a singly-linked list using CAS.
This implementation uses a cursor that points to three consecutive nodes in the list.
If the part of the list that the cursor is associated with is changed, the cursor becomes invalidated. 
To restore the validity of its own cursor, a process may have to perform CAS steps to help complete other processes' updates. 

Greenwald \cite{DLL-DCAS} presented a doubly-linked list implementation using 2-CAS.
In his approach, only one operation can make progress at a time.
Attiya and Hillel \cite{DLL-color} proposed a doubly-linked list implementation using 2-CAS. 
It has the nice property that only concurrent operations can interfere with one another only if they are changing  nodes close to each other. 
If there is no interference, an operation performs 13-15 CAS steps (and one 2-CAS). 
To avoid the ABA problem, a single word must store both a pointer and a counter.  
Their implementation does not update invalid cursors, so deletions might make other processes lose their place in the list. 
They also give a restricted implementation using single-word CAS, in which deletions can be performed only at the ends of the list.

Sundell and Tsigas \cite{DLL-CAS} gave the first non-blocking 
doubly-linked list using single-word CAS (although a word must store a bit and a pointer). 
Linearizable data structures are notoriously difficult to design, so detailed correctness proofs are essential.
In \cite{DLL-CAS}, a proof of the non-blocking property is provided, but to justify the claim of linearizability, the linearization points of operations are defined without providing a proof that they are correct.
In fact, their implementation appears to have minor errors:
using the Java PathFinder model checker \cite{JPF}, we discovered an execution that incorrectly dereferences a null pointer.
Their implementation is ingenious but quite complicated.
In particular, their helping mechanism is very complex, partly because operations can terminate before completing the necessary changes to the list,
so operations may have to help non-concurrent updates.
In our implementation, an update  helps only updates that are concurrent with itself,
 and moves do not help at all.
In the best case, their updates perform 2 to 4 CAS steps.
However, moves perform CAS steps to help complete updates.
In fact,  a series of deletions can construct long chains of deleted nodes whose pointers to adjacent nodes do not get updated by the deletions.
Then, a move operation may have to traverse this chain, performing CAS steps at every node.
As in \cite{DLL-CAS}, each update of our implementation appears to take effect at the first CAS.
When another
process deletes the item a cursor points to, we use a rather different approach from \cite{DLL-CAS} for recovering the location of the cursor using only reads of shared memory.





\mysection{The Sequential Specification} \label{ss-sec}
A list is a pair $(L, S)$ where
$L$ is a finite sequence of distinct items ending with a special end-of-list marker (EOL), and $S$ is a set of cursors.
The state of the list is initially $(\langle EOL \rangle, \emptyset)$.
Eight types of operations are supported: createCursor, destroyCursor, resetCursor, insertBefore, delete,  get, moveRight and moveLeft.
Each item $x$ in $L$ has a value denoted $x.val$, and values need not be distinct.

A cursor is a tuple $(name, item, invDel, invIns, id)$ that includes a unique name, the item in $L$ that the cursor is associated with, 
two boolean values that indicate whether the cursor is invalid for different operations (explained in more detail below)
and the id of the process that created the cursor.

A createCursor() creates a new cursor whose item is the first item in $L$ (which is EOL if $L$ contains only EOL) and destroyCursor($c$) destroys the cursor $c$.
A process $p$ can call an operation with a cursor $c$ only if $p$ itself created $c$ and $c$ has not been destroyed.
A resetCursor($c$) sets $c.item$ to the first item in $L$.
A get($c$) does not change $(L, S)$ and returns the $val$ field of $c.item$.
Move operations do not change $L$.
If $c.item \ne$ EOL, moveRight($c$) sets $c.item$ to the next item in $L$ and returns true;
otherwise, it does not change $(L, S)$ and returns false.
If $c.item$ is not the first item in $L$, moveLeft($c$) sets $c.item$ to the previous item in $L$ and returns true;
otherwise, it does not change $(L, S)$ and returns false.

Suppose a process $p$ has a cursor $c$ whose item is $x$.
If $x$ is deleted by another process $p'$, $c.item$ is set to the next item $y$ in $L$ and $c$ becomes invalid (i.e., $c.invDel$ becomes true).
Thus, the deletion cannot cause $c$ to lose its place in $L$.
Since $x$ is removed by $p'$, $p$ does not yet know that $c$ is no longer associated with $x$.
If $p$ then calls a delete operation with $c$ to attempt to remove $x$, it should not remove $y$.
To avoid this situation, the deletion sets $c.invDel$ to true.
When $c.invDel$ is true, the next operation that is called using it returns invalidCursor to indicate that the cursor has been moved. 
When an operation returns invalidCursor, the cursor's $invDel$ is set to false, making it valid again.

Suppose we wish to maintain $L$ so that values of items are sorted
and process $p$ has a cursor $c$ whose item's value is $5$.
Then, $p$ advances $c$ to the next item in the sequence, which has value $8$.
If $7$ is inserted by another process $p$ before $8$, $c$ becomes invalid for insertion (i.e., $c.invIns$ becomes true).
This invalidation ensures that an item can be inserted between two specific items in the list.
Since $7$ is inserted by $p'$, $p$ does not yet know that the item before $8$ is $7$.
If $p$ then calls an insertBefore operation with $c$ to attempt to insert $6$ before $8$, it should not succeed because that would place $6$ between $7$ and $8$.
Thus, when $c.invIns$ is true and the next operation called with $c$ is an insertBefore operation, it returns invalidCursor to indicate that a new item has been inserted before the cursor. 
When $c.invIns$ is true, the next operation called with $c$ sets $c.invIns$ to false again.

A more formal sequential specification is given in Appendix \ref{formal-spec}.


\mysection{The Non-blocking Implementation}

\begin{figure}[t]
\vspace{-10pt}
\centering
\includegraphics[width=\textwidth]{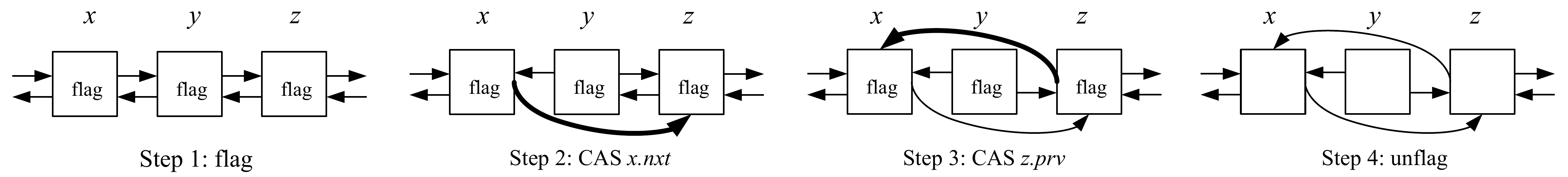}
\vspace{-25pt}
\caption{delete}
\label{del-fig}
\end{figure}
\begin{figure}[t]
\vspace{-8pt}
\centering
\includegraphics[width=\textwidth]{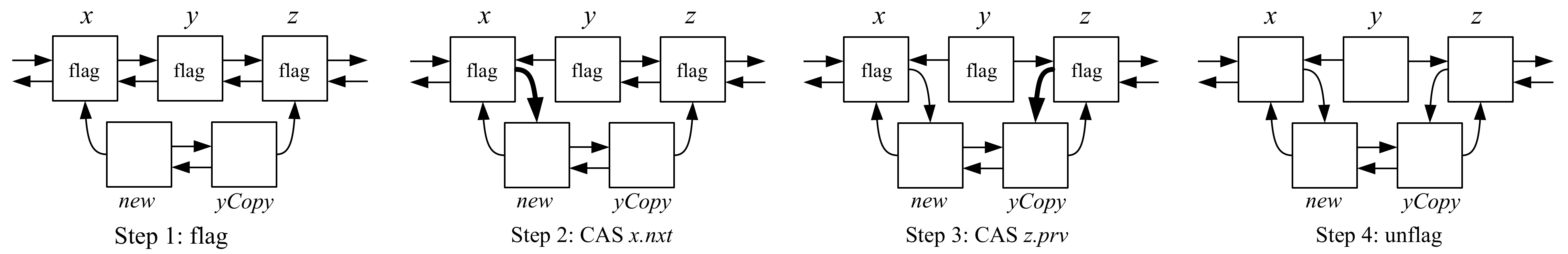}
\vspace{-25pt}
\caption{insertBefore}
\label{ins-fig}
\end{figure}

List items  are represented by Node objects, which have pointers to adjacent Nodes.
A cursor is represented in a process's local memory by a single pointer to a Node.
Updates are done in several steps as shown in Fig.~\ref{del-fig} and \ref{ins-fig}.
To avoid simultaneous updates to overlapping parts of the list, an update
flags a Node before removing it or changing one of its pointers.
A Node is flagged by storing a pointer to an Info object, which is a descriptor of the update, so that other updates
can help complete it. 
List pointers are updated using CAS
so that helpers cannot perform an operation more than once.

The correctness of algorithms using CAS often depends on the fact that, if a CAS on variable $V$ succeeds, $V$ has not changed since an earlier read. 
An ABA problem occurs when $V$ changes from one value to another and back before the CAS occurs, causing the CAS to succeed when it should not.  
When a Node $new$ is inserted between Node $x$ and $y$, we replace $y$ by a new copy, $yCopy$ (Fig.~\ref{ins-fig}). 
This avoids an ABA problem that would occur if, instead,
insertBefore simply 
changed the pointers in $x$ and $y$ to {\it new}, because
a subsequent deletion of {\it new} could then change $x$'s pointer back to $y$ again. 
Creating a new copy of $y$ 
also makes invalidation of Cursors for insertions easy.
An insertion of a Node before $y$ writes a permanent pointer to $yCopy$ in $y$ before replacing $y$, 
so that any other process whose Cursor is at $y$ can detect that an insertion has occurred there and update its Cursor to~$yCopy$.

The objects used in our implementation are described in 
line \ref{type-def-begin} to \ref{type-def-end} of Fig.~\ref{code-fig}.
A Node has the following fields.
The {\it val} field contains the item's value, 
{\it nxt} and {\it prv} point to the next and previous Nodes in the list, 
{\it copy} points to a new copy of the Node (if any),
{\it info} points to an Info object that is the descriptor of the update that last flagged the Node, 
and
{\it state} is initially ordinary and is set to copied (before the Node is replaced by a new copy) or marked (before the Node is deleted).
The {\it info} field is initially set to a dummy Info object, {\it dum}. 
The {\it info}, {\it nxt} and {\it prv} fields of a Node are changed using CAS steps.
We call the steps that try to modify these three fields {\it flag CAS}, {\it forward CAS} and {\it backward CAS} steps, respectively.
To avoid special cases, we add sentinel Nodes $head$ and $tail$, which do not contain values, at the ends of the list.
They are never changed and Cursors never move to $head$ or $tail$.
The last Node before $tail$ always contains the value EOL. 

Info objects are used as our operation descriptors.  An Info object $I$ has the following fields, which do not change after $I$ is created.
$I.nodes[0..2]$ stores the three Nodes $x$, $y$, $z$ to be flagged before changing the list.
$I.oldInfo[0..2]$  stores the expected values to be used by the flag CAS steps on $x,y$ and $z$. 
$I.newNxt$ and $I.newPrv$  store the new values for the forward and backward CAS steps on {\it x.nxt} and {\it z.prv}. 
$I.rmv$ indicates whether $y$ should be deleted from the list or replaced by a new copy.
$I.status$, indicates whether the update is inProgress (the initial value), committed (after the update is completed) or aborted (after a node is not flagged successfully). 
(One exception is the dummy Info object {\it dum} whose $status$ is initially aborted.)
A Node is {\it flagged for $I$} if its {\it info} field is $I$ and $I.status=$ inProgress.
Thus, setting $I.status$ to committed or aborted also has the effect of removing $I$'s flags.
As with locks, successful flagging of the three nodes guarantees that the operation will be completed successfully without interference from other operations. 
Unlike locks, if the process performing an update crashes after flagging, other processes may complete its update using the information in $I$.
An update attempts to flag a Node $v$ using a CAS step on $v.info$, which fails
if the Node is already flagged by another concurrent update; in this case, the operation is retried after helping the other update. 

\vspace*{-4mm}
\paragraph{Detailed Description of the Algorithms}
Pseudo-code for our implementation is given in Fig.~\ref{code-fig}.

Since a Cursor $c$ is a pointer in a process's local memory, it becomes out of date if the Node it points to is deleted or replaced by another process's update.
Thus, at the beginning of an update, move or get operation called with $c$, {\bf updateCursor}($c$) is called to bring $c.node$ up to date. 
If {\it c.node} has been replaced with a new copy by an insertBefore, 
updateCursor follows the copy pointer (line \ref{updateCursor-set-new-copy})
and sets $invIns$ to true (line \ref{updateCursor-set-invalidIns}). 
Similarly, if {\it c.node} has been deleted, 
updateCursor follows the {\it nxt} pointer (line \ref{updateCursor-set-nxt}), which is the next Node at the time of deletion,
and sets
{\it invDel} to true  (line~\ref{updateCursor-set-invalid}).
UpdateCursor repeats the loop at line \ref{updateCursor-check}--\ref{updateCursor-end-loop} until the test on line \ref{updateCursor-check} indicates that {\it c.node} is in the~list.




\begin{figure}
\begin{minipage}{0.5 \textwidth}
{\footnotesize 
\begin{enumerate}
{\sffamily
\addtolength{\itemsep}{-0.42\baselineskip}
\item {\bf type Cursor} \label{type-def-begin}
\item \tab Node $node$ \hfill $\rhd$ location of Cursor\\
 \vspace*{-1mm}
\item {\bf type Node}
\item \tab Value $val$ 
\item \tab Node {\it nxt} \hfill $\rhd$ next Node
\item \tab Node {\it prv} \hfill $\rhd$ previous Node 
\item \tab Node $copy$ \hfill $\rhd$ new copy of Node (if any)
\item \tab Info $info$ \hfill $\rhd$ descriptor of update
\item \tab \{copied, marked, ordinary\} {\it state} \\ \hspace*{0pt} \hfill $\rhd$ shows if Node is replaced or deleted \\
\vspace*{-1mm}
\item {\bf type Info} 
\item \tab Node[3] $nodes$ \hfill $\rhd$ Nodes to be flagged
\item \tab Info[3] $oldInfo$ \hfill $\rhd$ expected values of CASs that flag
\item \tab Node $newNxt$ \hfill $\rhd$ set $nodes[0].nxt$ to this
\item \tab Node $newPrv$ \hfill $\rhd$ set $nodes[2].prv$ to this
\item \tab Boolean $rmv$ \hfill $\rhd$ is $I.nodes[1]$ being deleted?
\item \tab \{inProgress, committed, aborted\} {\it status}\\  \label{type-def-end}
\vspace*{-1mm}
\item \makebox{\shortstack[l]{{\bf insertBefore}($c$: Cursor, $v$: Value):\{true, invalidCursor\}}}
\item \tab while(true) \label{ins-start-loop}
\item \twotab $\langle y, yInfo, z, x, invDel, invIns \rangle \leftarrow$ \\ \fourtab {\bf updateCursor}($c$)\label{ins-call-updateCursor}
\item \twotab if $invDel$ or $invIns$ then return invalidCursor \label{ins-return-invalid}
\item \twotab $nodes \leftarrow [x, y, z]$ \label{ins-set-nodes}
\item \twotab $oldInfo \leftarrow [x.info, yInfo, z.info]$ \label{ins-read-info}

\item \twotab if {\bf checkInfo}($nodes$, $oldInfo$) then \label{ins-call-checkInfo}
\item \threetab $new \leftarrow$ new Node($v$, null, $x$, null, $dum$, ordinary)\label{ins-create-newNode}
\item \threetab $yCopy \leftarrow$ new Node($y.val$, $z$, $new$, null, $dum$, \\ \fourtab ordinary)\label{ins-create-nodeCopy}
\item \threetab $new.nxt \leftarrow yCopy$ \label{ins-set-newNode}

\item \threetab $I \leftarrow$  new Info($nodes$, $oldInfo$, $new$, $yCopy$, \\ \fourtab false, inProgress)\label{ins-create-info}
\item \threetab if {\bf help}($I$) then \label{ins-call-help}
\item \fourtab $c.node \leftarrow yCopy$ \label{ins-set-cursor}
\item \fourtab return true \label{ins-return-true} \\
 \vspace*{-1mm}

\item{\bf delete}($c$: Cursor):\{true, false, invalidCursor\}
\item \tab while(true) \label{del-start-loop}

\item \twotab $\langle y, yInfo, z, x, invDel,$ -$\rangle \leftarrow$ {\bf updateCursor}($c$) \label{del-call-updateCursor} 
\item \twotab if $invDel$ then return invalidCursor \label{del-return-invalid}
\item \twotab $nodes \leftarrow [x, y, z]$ \label{del-set-nodes}
\item \twotab $oldInfo \leftarrow [x.info, yInfo, z.info]$ \label{del-read-info}

\item \twotab if {\bf checkInfo}($nodes$, $oldInfo$) then \label{del-call-checkInfo}
\item \threetab if $y.val =$ EOL then return false \label{del-return-false}
\item \threetab $I \leftarrow$ new Info($nodes$, $oldInfo$, $z$, $x$, true,  \\ \fourtab inProgress)\label{del-create-info}
\item \threetab if {\bf help}($I$) then \label{del-call-help}
\item \fourtab $c.node \leftarrow z$ \label{del-set-cursor}
\item \fourtab return true \label{del-return-true} \\
 \vspace*{-1mm}

\item{\bf moveLeft}($c$: Cursor):\{true, false, invalidCursor\}
\item \tab \makebox{\shortstack[l]{$\langle y, -, -, x, invDel,$ -$\rangle \leftarrow$  {\bf updateCursor}($c$)}}\label{moveLeft-call-updateCursor} 
\item \tab if $invDel$ then return invalidCursor  \label{moveLeft-return-invalid}
\item \tab if $x = head$ then return false \label{moveLeft-return-false1}

\item \tab if $x.state \ne$ ordinary and $x.prv.nxt \ne x$ and  \\ \twotab $x.nxt = y$ then \label{moveLeft-check-prvNode} 
\item \twotab if $x.state =$ copied then \label{moveLeft-check-state}
\item \threetab $c.node \leftarrow x.copy$ \label{moveLeft-set-cursor-copy}
\item \twotab else
\item \threetab $w \leftarrow x.prv$ \label{moveLeft-read-prvPrvNode}
\item \threetab if $w = head$ then return false \label{moveLeft-return-false2}
\item \threetab $c.node \leftarrow w$ \label{moveLeft-set-cursor-remove} 
\item \tab else $c.node \leftarrow x$ \label{moveLeft-set-cursor}
\item \tab return true \\
 \vspace*{-1mm}

\newcounter{c}
\setcounter{c}{\value{enumi}}
}
\end{enumerate}
}
\end{minipage}
\begin{minipage}{0.5 \textwidth}
{\footnotesize
\begin{enumerate}
\setcounter{enumi}{\value{c}}
{\sffamily
\addtolength{\itemsep}{-0.42\baselineskip}
\item {\bf moveRight}($c$: Cursor):\{true, false, invalidCursor\}
\item \tab \makebox{\shortstack[l]{$\langle y, -, z, -, invDel,$ -$\rangle \leftarrow$  {\bf updateCursor}($c$)}}\label{moveRight-call-updateCursor} 
\item \tab if $invDel$ then return invalidCursor \label{moveRight-return-invalid}
\item \tab if $y.val =$ EOL then return false \label{moveRight-return-false}
\item \tab $c.node \leftarrow z$ \label{moveRight-set-cursor} 
\item \tab return true \label{moveRight-return-true} \\
 \vspace*{-1.5mm}
 
\item {\bf createCursor}():Cursor
\item \tab return new Cursor($head.nxt$) \label{create-cursor} \label{createCursor-set-cursor}\\
\vspace*{-1.5mm}

\item {\bf destroyCursor}($c$: Cursor)
\item \tab return ack\\
\vspace*{-1.5mm}

\item{\bf resetCursor}($c$: Cursor)
\item \tab $c.node \gets head.nxt$ \\ \label{resetCursor-set-cursor} 
 \vspace*{-1.5mm}

\item{\bf get}($c$: Cursor):Value
\item \tab $\langle y, -, -, -, invDel,$ -$\rangle \leftarrow$  {\bf updateCursor}($c$)\label{get-call-updateCursor} 
\item \tab if $invDel$ then return invalidCursor\label{get-return-invalid}
\item \tab return $y.val$ \\
 \vspace*{-1.5mm}

\item{\bf updateCursor}($c$: Cursor):$\langle$Node, Info, Node, Node, \\ \twotab Boolean, Boolean$\rangle$
\item \tab $invDel \leftarrow$ false \label{updateCursor-initial-invalid}
\item \tab $invIns \leftarrow$ false \label{updateCursor-initial-invalidIns}
\item \tab while($c.node.state \ne$ ordinary and \\ \threetab $c.node.prv.nxt \ne c.node$)\label{updateCursor-check}
\item \twotab if $c.node.state =$ copied then \hfill $\rhd$ $node$ replaced\label{updateCursor-check-state}
\item \threetab $c.node \leftarrow c.node.copy$ \label{updateCursor-set-new-copy}
\item \threetab $invIns \leftarrow$ true \label{updateCursor-set-invalidIns}
\item \twotab else \hfill $\rhd$ $node$ deleted\label{updateCursor-check-mark}
\item \threetab $c.node \leftarrow c.node.nxt$ \label{updateCursor-set-nxt}
\item \threetab $invDel \leftarrow$ true \label{updateCursor-set-invalid} \label{updateCursor-end-loop}

\item \tab $info \leftarrow  c.node.info$ \label{updateCursor-read-info}
\item \tab return $\langle c.node, info, c.node.nxt, c.node.prv, invDel$, \\ \threetab $invIns\rangle$ \\ \label{updateCursor-return} \label{updateCursor-read-ptr}
 \vspace*{-2mm}

\item{\bf checkInfo}($nodes$: Node[3], $oldInfo$: Info[3]):Boolean 
\item \tab for $i \leftarrow 0$ to $2$,
\item \twotab if $oldInfo[i].status =$ inProgress then \label{checkInfo-check-inProgress}
\item \threetab {\bf help}($oldInfo[i]$) \label{checkInfo-call-help}
\item \threetab  return false \hfill $\rhd$ in progress update on $nodes[i]$\label{checkInfo-return-false1}

\item \tab for $i \leftarrow 0$ to $2$,
\item \twotab if $nodes[i].state \ne$ ordinary then \label{checkInfo-check-state} 
\item \threetab return false \hfill $\rhd$ $nodes[i]$ removed\label{checkInfo-return-false2}

\item \tab for $i \leftarrow 1$ to $2$,
\item \twotab if $nodes[i].info \ne oldInfo[i]$ then return false\label{checkInfo-return-false3} 

\item \tab return true \\
 \vspace*{-1.5mm}

\item{\bf help}($I$: Info):Boolean
\item \tab $doPtrCAS \leftarrow$ true
\item \tab $i \leftarrow 0$

\item \tab while ($i < 3$ and $doPtrCAS$) \label{help-check-before-nodes}
\item \twotab \makebox{\shortstack[l]{CAS($I.nodes[i].info$, $I.oldInfo[i]$, $I$) $\rhd$ {\bf flag CAS}}}\label{help-flag}
\item \twotab $doPtrCAS \leftarrow (I.nodes[i].info = I)$ \label{help-set-doPtrCAS}
\item \twotab $i \leftarrow i + 1$ 

\item \tab if $doPtrCAS$ then \label{help-check-doPtrCAS}
\item \twotab if $I.rmv$ then $I.nodes[1].state \leftarrow$ marked \label{help-set-marked}
\item \twotab else 
\item \threetab $I.nodes[1].copy \leftarrow I.newPrv$ \label{help-set-copy}
\item \threetab $I.nodes[1].state \leftarrow$ copied \label{help-set-copied}

\item \twotab CAS($I.nodes[0].nxt$, $I.nodes[1]$, $I.newNxt$) \\   \vspace*{-.3mm} \hspace*{0pt} \hfill $\rhd$ {\bf forward CAS}\label{help-forward-CAS}
 \vspace*{-.1mm}

\item \twotab CAS($I.nodes[2].prv$, $I.nodes[1]$, $I.newPrv$) \\  \hspace*{0pt} \hfill $\rhd$ {\bf backward CAS}\label{help-backward-CAS}
 \vspace*{-.5mm}
\item \twotab $I.status \leftarrow$ committed \label{help-set-committed}

\item \tab else if $I.status =$ inProgress then $I.status \leftarrow$ aborted\label{help-set-aborted}\label{help-check-inProgress}

\item \tab return ($I.status =$ committed) \label{help-return-false}\label{help-check-status} \label{help-return-true}
 \vspace*{-2mm}

\setcounter{c}{\value{enumi}}
}
\end{enumerate}
}
\end{minipage}
\caption{Pseudo-code for a non-blocking doubly-linked list.\label{code-fig}}
\end{figure}



After calling updateCursor, each update $op$ calls {\bf checkInfo} to see if some Node that $op$ wants to flag is flagged with an Info object $I'$ of another update.
If so, it calls help($I'$) (line \ref{checkInfo-call-help}) to try completing the other update, 
and returns false to indicate $op$ should retry.
Similarly, if checkInfo sees that 
one of the Nodes is already removed from the list (line \ref{checkInfo-check-state}), it returns false, causing $op$ to retry.
If checkInfo sees that the $info$ of $y$ or $z$ has  already been changed by another process (line \ref{checkInfo-return-false3}), to avoid flagging $x$, it returns false, causing $op$ to retry.
If checkInfo returns true, $op$ creates a new Info object $I$ 
for its update (line \ref{ins-create-info} or \ref{del-create-info}) and 
calls help($I$) to try to complete its own update (line \ref{ins-call-help} or~\ref{del-call-help}).


The {\bf help}($I$) routine 
performs the real work of the update.
First, it uses flag CAS steps to store 
$I$ in the {\it info} fields of the Nodes to be flagged (line \ref{help-flag}).
If help($I$) sees a Node $v$ is not flagged successfully (line \ref{help-set-doPtrCAS}), help($I$) checks if {\it I.status} is inProgress (line \ref{help-check-inProgress}).  If so, 
it follows that no helper of $I$ succeeded in flagging all three  nodes; otherwise $I$'s 
flag on $v$ could not have been removed while $I$ is inProgress.
So, $v$ was flagged by another update before help($I$)'s flag CAS.
Thus, {\it I.status} is set to aborted (line \ref{help-set-aborted}) and help($I$) returns false (line \ref{help-return-false}), causing $op$ to retry.

If the Nodes $x,y$ and $z$ in $I.nodes$ 
are all flagged successfully with $I$,  {\it y.state} is set to marked 
(line \ref{help-set-marked}) for a deletion, or copied (line \ref{help-set-copied}) 
for an insertion.
In the latter case, {\it y.copy} is first set to the new copy (line \ref{help-set-copy}). 
Then, 
a  forward CAS  (line \ref{help-forward-CAS}) changes {\it x.nxt} and 
 a backward CAS  (line \ref{help-backward-CAS}) changes {\it z.prv}.
Finally, help($I$) sets $I.status$ to committed (line \ref{help-set-committed}) and returns true (line \ref{help-return-true}).
A {\it CAS of $I$} refers to a CAS step executed inside help($I$).
We prove below that 
the {\it first} forward 
and {\it first} backward CAS of $I$ among all calls to help($I$) succeed (and no others do).


We say a Node $v$ is {\it reachable} 
if there is a path of {\it nxt} pointers from $head$ to $v$. 
At all times, the reachable Nodes correspond to the items in the list.
So, the update that created $I$ is linearized at the first forward CAS of $I$.
Just after this CAS, $y$ becomes unreachable (step 2 of Fig.~\ref{del-fig} and \ref{ins-fig}).
We prove that no process changes {\it y.nxt} or {\it y.prv} after that, 
so $y.prv$ remains equal to $x$.
Since there is no ABA problem, {\it x.nxt} is never set back to $y$ after $y$ becomes unreachable.
Thus, the test $y.prv.nxt \ne y$ tells us whether $y$ has become unreachable.
(After $y$ becomes unreachable we also have $y.state\neq$ ordinary.)

Both the {\bf insertBefore}($c,v$) and {\bf delete}($c$) operations have the same structure.
They first call updateCursor($c$) 
to bring the Cursor $c$ up to date, and return invalidCursor if this routine indicates $c$ has been invalidated.
Then, they call checkInfo 
to see if there is interference by other updates.  If not, they create an Info object $I$ and call help($I$) to complete the update.  If unsuccessful, they retry.

\ignore{
The {\bf insertBefore}($c$, $v$) first calls updateCursor($c$) (line \ref{ins-call-updateCursor}).
Let $\langle y$, -, -, $x$, $invDel$, $invIns \rangle$ be the result returned by updateCursor.
If $invDel$ or $invIns$ is true, the insertBefore returns invalidCursor since $c$ is invalidated by a delete or insertBefore (line \ref{ins-return-invalid}).
Otherwise, the insertion assures that there is no interference with other concurrent processes by calling checkInfo (line \ref{ins-call-checkInfo}).
Then, it creates Node $new$ whose $val$ is $v$ (line \ref{ins-create-newNode}) and a new copy of $y$ (line \ref{ins-create-nodeCopy}).
After that, the operation creates an Info object (line \ref{ins-create-info}) and calls the help routine (line \ref{ins-call-help}) to replace $y$ by the new copy and insert $new$ between $x$ and the new copy of $y$.

The {\bf delete($c$)} first calls updateCursor($c$) (line \ref{del-call-updateCursor}).
Let $\langle y$, -, -, -, -, -$\rangle$ be the result returned by updateCursor.
Then, the delete assures that there is no interference with other processes by calling checkInfo (line \ref{del-call-checkInfo}).
After that, the operation creates an Info object (line \ref{del-create-info}) and calls the help routine (line \ref{del-call-help}) to remove $y$ from the list. 
}

A {\bf moveRight($c$)} calls updateCursor($c$) (line \ref{moveRight-call-updateCursor}), which sets $c.node$ to a Node $y$ and also returns a Node $z$ read from $y.nxt$.
We show there is a time during move when $y$ is reachable and $y.nxt=z$.
If $y.val =$ EOL, 
the operation cannot move $c$ and returns false.
Else, 
 it sets $c.node$ to $z$ (line \ref{moveRight-set-cursor}).


A {\bf moveLeft($c$)} is more complex because {\it prv} pointers are updated {\it after} an update's linearization point, so they are sometimes inconsistent with the true state of the list. 
A moveLeft first calls updateCursor($c$) (line \ref{moveLeft-call-updateCursor}),
which updates $c.node$ to some Node $y$ and also returns a Node $x$ read from $y.prv$.
If $x$ is $head$ (line \ref{moveLeft-return-false1}), the operation cannot move $c$ to $head$ and returns false.
If the test on line  \ref{moveLeft-check-prvNode} indicates $x$ was reachable, $c.node$ is set to $x$ (line \ref{moveLeft-set-cursor}).
This is also done if $x.nxt\neq y$; in this case, we can show that $y$ became unreachable
during the move operation, but $x.nxt$ pointed to $y$ just before it became unreachable.
Otherwise, $x$ has become unreachable and the test $x.nxt=y$ on line \ref{moveLeft-check-prvNode} ensures that $x$ was the element before $y$ when it became unreachable.  
If $x$ was replaced by an insertion, $c.node$ is set
to that replacement node (line \ref{moveLeft-set-cursor-copy}).  If $x$ was removed by a
deletion, we set $c.node$ to $x.prv$ (line \ref{moveLeft-set-cursor-remove}), unless
that node is {\it head}.
We prove in Lemma \ref{ml-lin}, below, that whenever moveLeft updates $c.node$ to some value $v$,
there is a time during the operation when $v$ is reachable and $v.nxt=y$.

\mysection{Correctness Proof}
The detailed proof of correctness (available in \cite{tech-rep})
is quite lengthy, so we give a brief sketch in three parts.
An execution 
is a sequence of configurations, $C_0, C_1, ...$ such that, for each $i \ge 0$, $C_{i+1}$ follows from $C_i$ by a step of the implementation.
For the proof, we assign each Node $v$ a positive real value, called its {\it abstract value}, denoted $v.absVal$.
The $absVal$ of {\it head}, EOL and {\it tail} are 0, 1 and 2 respectively.
When insertBefore creates the Nodes $new$ and $yCopy$ (see Fig.~\ref{ins-fig}), $yCopy.absVal =y.absVal$ and $new.absVal = (x.absVal+y.absVal)/{2}$.
The following basic facts are easy to prove. 
\begin{invariant}
\label{inv}
- Any field that is read in the pseudo-code is non-null.\\
\hspace*{.88in} - Cursors do not point to {\it head} or {\it tail}.\\ 
\hspace*{.88in} - If $v.nxt=tail$, then $v.val=$ EOL.\\
\hspace*{.88in} - If $v.nxt=w$ or $w.prv=v$ then $v.absVal<w.absVal$.
\end{invariant}

\vspace{-12pt}
\paragraph{Part 1: Flagging}
Part 1 proves $v$ is flagged for $I$ when the first forward CAS or first backward CAS of an Info object $I$ is applied to Node $v$.  
We first show there is no ABA problem on {\it info} fields.
\begin{lemma}
\label{infoABA}
The $info$ field of a Node is never set to a value that has been stored there previously.
\end{lemma}
\noindent {\it Proof sketch.} 
The old value used for $I$'s flag CAS on Node $v$ was read from the {\it info} field of $v$ before $I$ is created.
So, every time {\it v.info} is changed from $I'$ to $I$, $I$ is a newer Info object than $I'$.
\qed

By Lemma \ref{infoABA}, only the first flag CAS of $I$ on each Node in {\it I.nodes} can succeed since all such CAS steps use the same expected value.
We say $I$ is {\it successful} if these three first flag CAS steps all succeed. 
\begin{lemma}
\label{status-lem}
After $v.info$ is set to $I$, it remains $I$ until $I.status\ne$ inProgress.
\end{lemma}
\noindent {\it Proof sketch.} 
If {\it v.info} is changed from $I$ to $I'$, a call to checkInfo on line \ref{ins-call-checkInfo} or \ref{del-call-checkInfo} must have seen that $I.status\ne$ inProgress 
before $I'$ was created at line \ref{ins-create-info} or \ref{del-create-info}.
\qed
\begin{observation}
\label{succ-info-obs}
If any process executes line \ref{help-set-marked}--\ref{help-set-committed} inside help($I$), then $I$ is already successful.
\end{observation}
\begin{lemma}
\label{succ-info-lem}
If $I$ is successful, $I.status$ is never aborted.
Otherwise, $I.status$ is never committed.
\end{lemma}
\noindent {\it Proof sketch.} 
If $I$ is not successful, the claim follows from Observation \ref{succ-info-obs}.
If $I$ is successful, the first flag CAS on each Node in $I.nodes$ succeeds.
By Lemma \ref{status-lem}, every call to help($I$) evaluates the test on line \ref{help-check-doPtrCAS} to true until $I.status\ne$ inProgress.
So, no process sets $I.status$ to aborted on line \ref{help-set-aborted}. 
\qed
\begin{lemma}
\label{flag-ptr}
For each of lines \ref{help-set-marked}--\ref{help-set-committed},
when the first execution of that line among all calls to help($I$) occurs, all Nodes in {\it I.nodes} are flagged for $I$. 
\end{lemma}
\noindent {\it Proof sketch.}
Suppose one of lines  \ref{help-set-marked}--\ref{help-set-committed} is executed inside help($I$).
By Observation \ref{succ-info-obs}, a flag CAS of $I$ already succeeded on each Node in $I$.{\it nodes}.
By Lemma \ref{succ-info-lem}, $I.status$ is never aborted.
By Lemma \ref{status-lem}, 
all three Nodes remain flagged for $I$ until some help($I$) sets $I.status$ to committed on line~\ref{help-set-committed}.
\qed

\vspace{-12pt}
\paragraph{Part 2: Forward and Backward CAS Steps}
Let $\langle y_I$, -, $z_I$, $x_I$, -, -$\rangle$ be the result updateCursor($c$) returns on line \ref{ins-call-updateCursor} or \ref{del-call-updateCursor} before creating $I$ on line \ref{ins-create-info} or \ref{del-create-info}. 
Part 2 of our proof shows that successful flagging ensures that $x_I$, $y_I$ and $z_I$ are three consecutive Nodes in the list just before the first forward CAS of $I$, 
and that the first forward and the first backward CAS of $I$ succeed (and no others do). 
\begin{lemma}
\label{marked}
At all configurations after $I$ becomes successful, $y_I.info=I$.
\end{lemma}
\noindent {\it Proof sketch.} 
To derive a contradiction, assume $y_I.info$ is changed from $I$ to $I'$.
Before creating $I'$, the call to checkInfo returns true, so it sees $I.status\ne$ inProgress at line \ref{checkInfo-check-inProgress} and then $y_I.state=$ ordinary at line \ref{checkInfo-check-state}.
This contradicts the fact that before {\it I.status} is set to committed at line \ref{help-set-committed}, $y_I.state$ is set to a non-ordinary value at line \ref{help-set-marked} or \ref{help-set-copied} 
(and is never changed back to ordinary).
\qed

\begin{lemma} \label{ptr-lem}
\begin{enumerate}
\addtolength{\itemsep}{-0.7\baselineskip}
\item The first forward and the first backward CAS of $I$ succeed and all other forward and backward CAS steps of $I$ fail.\label{first-ptr}
\item The $nxt$ or $prv$ field of a Node is never set to a Node that has been stored there before.\label{ABA-ptr}
\item At the configuration $C$ before the first forward CAS of $I$, $x_I$, $y_I$ and $z_I$ are reachable, $x_I.nxt=y_I$, $y_I.prv=x_I$, $y_I.nxt=z_I$ and $z_I.prv=y_I$.\label{ptr-reach-lem}
\item At all configurations after the first forward CAS of $I$, $y_I.prv=x_I$ and $y_I.nxt=z_I$.\label{frozen-lem}
\end{enumerate}
\end{lemma}
\noindent {\it Proof sketch.}  We use induction on the length of the execution.
Statement \ref{first-ptr}:
By induction hypothesis~\ref{ptr-reach-lem}, the first forward CAS of $I$ succeeds, since $x_I.nxt=y_I$ just before it.
By induction hypothesis \ref{ABA-ptr}, no other forward CAS of $I$ succeeds. 
By induction hypothesis \ref{ptr-reach-lem}, $z_I.prv$ was $y_I$ at some time before the first backward CAS of $I$.
All backward CASes of $I$ use $y_I$ as the expected value of $z_I.prv$, so only the first can succeed (by induction hypothesis \ref{ABA-ptr}). 
By Lemma \ref{flag-ptr}, $z_I.info = I$ at the first forward and first backward CAS of $I$, 
and hence at all times between,
by Lemma \ref{infoABA}. 
By Lemma \ref{flag-ptr}, no backward CAS of any other Info object changes $z_I.prv$ during this time.
So, the first backward CAS of $I$ succeeds.

Statement \ref{ABA-ptr}:
Intuitively, when the {\it nxt} field changes from $v$ to another value, $v$ is thrown away and never used again.
(See Fig.~\ref{del-fig} and \ref{ins-fig}).
Suppose the first forward CAS of $I$ changes $x_I.nxt$.
If $I$ is created by an insertBefore, the CAS changes $x_I.nxt$ to a newly created Node.
If $I$ is created by a delete, $z_I.info = I$ at the first forward CAS of $I$, by Lemma \ref{flag-ptr}.
No forward CAS of another Info object $I'$ can change  $x_I.nxt$ from $z_I$ to another value earlier, since then $z_I.info$ would have to be $I'$ at the first forward CAS of $I$, by Lemma \ref{marked}.
The proof for {\it prv} fields is symmetric.

\begin{figure}[t]
\vspace{-11pt}
\centering
\includegraphics[width=\textwidth]{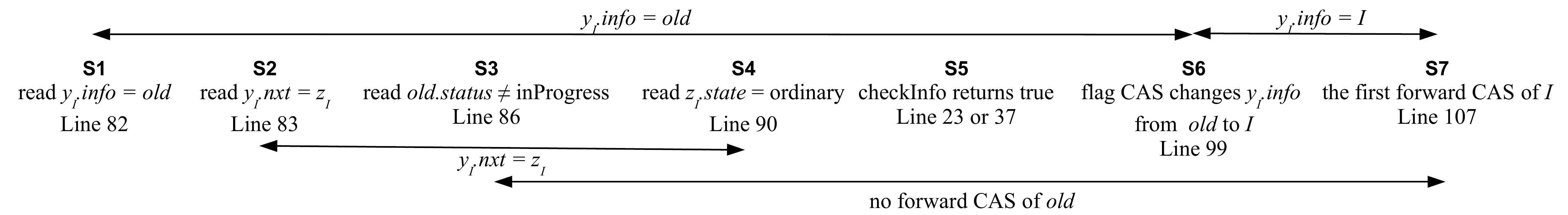}
\vspace{-24pt}
\caption{Sequence of events used in proof of Lemma \ref{ptr-lem}, Statement \ref{ptr-reach-lem}.}
\label{lem-fig}
\end{figure}

Statement \ref{ptr-reach-lem}:
First, we prove $y_I.nxt = z_I$ at $C$.
Before $I$ can be created, 
the sequence of steps $S1, ..., S5$ shown in Fig.~\ref{lem-fig} must occur.
By Lemma \ref{flag-ptr}, $y_I.info$ is set to $I$ by some step $S6$ and $y_I.info = I$ at $S7$.
By Lemma \ref{infoABA}, $y_I.info = old$ between $S1$ and $S6$ and $y_I.info = I$ between $S6$ and $S7$.
So, by Lemma \ref{flag-ptr}, only the first forward CAS of {\it old} can change $y_I.nxt$ between $S1$ and~$S7$.
Before $y_I.nxt$ can be changed from $z_I$ to another value by help({\it old}), $z_I.state$ is set to marked or copied (and it can never be changed back to ordinary).
So, $y_I.nxt$ is still $z_I$ at $S4$.
The first forward CAS of $old$ does not occur after $S3$ since {\it old.state} is already committed or aborted at $S3$.
So, $y_I.nxt$ is still $z_I$ at $C$.

By a similar argument, $y_I.prv=x_I$ and $x_I$, $y_I$ and $z_I$ are reachable in $C$.
The {\it prv} and {\it nxt} field of two adjacent  reachable Nodes might not be consistent at $C$ only if $C$ is between the first forward and first backward CAS of some Info object $I'$ and one of the two Nodes is flagged for $I'$ 
(step 2 of Fig.~\ref{del-fig} and \ref{ins-fig}).
Since $x_I$, $y_I$ and $z_I$ are flagged for $I$ at $C$ (by Lemma \ref{flag-ptr}), $x_I.nxt=y_I$ and $z_I.prv=y_I$ at $C$. 

Statement \ref{frozen-lem}:
By induction hypothesis \ref{ptr-reach-lem}, $y_I.prv=x_I$ at the first forward CAS of $I$.
By Lemma \ref{marked}, $y_I.info$ is always $I$ after that.
So, by Lemma \ref{flag-ptr}, no backward CAS of another Info object changes {\it y$_I$.prv} after the first forward CAS of $I$.
Similarly for {\it y$_I$.nxt $=z_I$}. 
\qed

Consider Fig.~\ref{del-fig} and \ref{ins-fig}.
By Lemma \ref{ptr-lem}.\ref{ptr-reach-lem}, just before the first forward CAS of $I$, the {\it nxt} and {\it prv} field of $x_I$, $y_I$ and $z_I$ are as shown in step 1.
By Lemma \ref{ptr-lem}.\ref{first-ptr}, this CAS changes $x_I.nxt$ as shown in step 2 and the first backward CAS of $I$ changes $z_I.prv$ as shown in step 3.
The next lemma follows easily.
\begin{lemma}
\label{reach-lem}
A Node $v$ that was reachable before is reachable now iff $v.state=$ ordinary or $v.prv.nxt=v$.
\end{lemma}

\vspace{-12pt}
\paragraph{Part 3: Linearizability}
Part 3 of our proof shows that operations are linearizable.
The following lemmas show that there is a linearization point for each move operation.
In the following four proofs, $\langle y$, -, $z$, $x$, -, -$\rangle$ denotes the result updateCursor($c$) returns on line \ref{moveLeft-call-updateCursor} or \ref{moveRight-call-updateCursor} and 
$C_{\ref{updateCursor-check}}$ be the configuration before the last execution of line \ref{updateCursor-check} inside that call to updateCursor.
\begin{lemma}
\label{mr-lin}
If moveRight($c$) changes $c.node$ from $y$ to $z$ at line \ref{moveRight-set-cursor}, there is a configuration during the move when $y.nxt = z$ and $y$ is reachable.
\end{lemma} 
\noindent {\it Proof sketch.}
It follows from Lemma \ref{reach-lem}, that $y$ is reachable in $C_{\ref{updateCursor-check}}$.
(Some reasoning is required to see this, since line \ref{updateCursor-check} does three reads of shared memory.)
If $y$ is reachable when $y.nxt=z$ on line \ref{updateCursor-read-ptr}, the claim is true then. 
Otherwise, $y$ became unreachable by a forward CAS of an Info object $I$ between $C_{\ref{updateCursor-check}}$ and line \ref{updateCursor-read-ptr} and $y = I.nodes[1]$.
By Lemma \ref{ptr-lem}.\ref{frozen-lem}, $y.nxt$ is always $I.nodes[2]$ after the CAS.
Since $y.nxt=z$ on line \ref{updateCursor-read-ptr}, $z = I.nodes[2]$ and, by Lemma \ref{ptr-lem}.\ref{ptr-reach-lem}, the claim is true just before the CAS. 
\qed
\begin{lemma}
\label{mr-false}
If moveRight($c$) returns false, there is a configuration during the move when $c.node.val =$ EOL and $c.node$ is reachable. 
\end{lemma} 
\noindent {\it Proof sketch.}
Lemma \ref{reach-lem} implies that $y$ is reachable in $C_{\ref{updateCursor-check}}$, so the lemma is true in $C_{\ref{updateCursor-check}}$.
\qed
\begin{lemma}
\label{ml-lin}
If moveLeft($c$) changes $c.node$ from $y$ to $v$ at line \ref{moveLeft-set-cursor-copy}, \ref{moveLeft-set-cursor-remove} or \ref{moveLeft-set-cursor}, 
there is a configuration during the move when $v.nxt = y$ and $v$ is reachable.
\end{lemma} 
\noindent {\it Proof sketch.}
Suppose $c.node$ is set on line \ref{moveLeft-set-cursor-remove}. 
Lemma \ref{reach-lem} implies $x$ is unreachable after line \ref{moveLeft-check-prvNode}.
By Lemma \ref{ptr-lem}.\ref{first-ptr}, $x$ became unreachable by the first forward CAS of an Info object $I$ and $x = I.nodes[1]$.
Since $c.node$ is set on line \ref{moveLeft-set-cursor-remove}, $x.state =$ marked on line \ref{moveLeft-check-state}, so $I$ is created by a delete.
By Lemma \ref{ptr-lem}.\ref{frozen-lem}, $x.nxt$ is always $I.nodes[2]$ after the forward CAS.
Since $x.nxt = y$ on line \ref{moveLeft-check-prvNode}, $y = I.nodes[2]$.
Since the read of $y.prv$ returns $x$ on line \ref{updateCursor-read-ptr}, the first backward CAS of $I$ did not occur before that read (step 2 of Fig.~\ref{del-fig}).
So, at some time during move, 
$I.nodes[0].nxt = I.nodes[2] = y$.
Since $x.prv$ is always $I.nodes[0]$ after the forward CAS of $I$,  the move sets $w$ to $I.nodes[0]$ on line \ref{moveLeft-read-prvPrvNode} and then sets $c.node$ to $w$.
So, at some time during move, 
$I.nodes[0].nxt = y$ and $I.nodes[0]$ is reachable, as required. 
For
line \ref{moveLeft-set-cursor-copy} and \ref{moveLeft-set-cursor}, the proof is similar to the case above and the proof of Lemma \ref{mr-lin}, respectively.~\qed
\begin{lemma}
\label{ml-false}
If moveLeft($c$) returns false, $c.node = head.nxt$ in some configuration during the move.
\end{lemma} 
\noindent {\it Proof sketch.}
If moveLeft returns on line \ref{moveLeft-return-false1}, the proof is similar to Lemma \ref{mr-lin}, since $x=head$ and
$c.node=head.nxt$ at some configuration during the move.
For line \ref{moveLeft-return-false2}, the proof is similar to Lemma \ref{ml-lin}, since $w=head$ and $c.node=head.nxt$ at some configuration during the move. 
\qed

Next, we define the linearization points.
Each move is linearized at the step after the configuration defined by  Lemma \ref{mr-lin}, \ref{mr-false}, \ref{ml-lin} or \ref{ml-false}.
If there is a forward CAS of an Info object created by an update, the update is linearized at the first such CAS.
Each createCursor and resetCursor is linearized at reading $head.nxt$.
Each get, each delete that returns false and each operation that returns invalidCursor is linearized 
at the first step of the last execution of line \ref{updateCursor-check} inside its last call to updateCursor.

We define $(\gL, \gS)$ to be an auxiliary variable of type list.
Each time an operation is linearized, the same operation is atomically applied to $(\gL, \gS)$ according to the sequential specification. 
Lemma \ref{c-lem} implies that each operation returns the same response as the corresponding operation on $(\gL, \gS)$.
The {\it absVal} of the item containing EOL is 1.
If an item $\gq$ is inserted between items $\gp$ and $\gr$, $\gq.absVal=(\gp.absVal+\gr.absVal)/{2}$.
If $\gq$ is inserted before the first item $\gr$, $\gq.absVal=\frac{\gr.absVal}{2}$.
In Lemma \ref{c-lem}, we use $absVal$ to show there is an one-to-one correspondence between Nodes in the list and items in $\gL$.

The cursor in $\gS$ corresponding to Cursor $c$ is denoted $\gc$. 
Since $c$ is a local variable, $c.node$ might become out of date when other processes update $\gc$.
The true location of a cursor whose $c.node$ is $x$ is 
\vspace{-10pt}
\[\mathrm{realNode}(x) = \left\{
 \begin{array}{l l l}
$realNode$(x.copy) & \quad \text{if $x.state =$ copied and $x$ is unreachable},\\
$realNode$(x.nxt) & \quad \text{if $x.state =$ marked and $x$ is unreachable},\\
x & \quad \text{otherwise}.
  \end{array}
\vspace{-10pt}
  \right. \]

\vspace*{-3mm}
\noindent
An update is {\it successful} if it is linearized at a forward CAS and 
a move is {\it successful} if it sets $c.node$ on line \ref{moveLeft-set-cursor-copy}, \ref{moveLeft-set-cursor-remove}, \ref{moveLeft-set-cursor} or \ref{moveRight-set-cursor}.
We prove Lemma \ref{c-lem} by induction on the length of the execution. 
\begin{lemma}
\label{c-lem}
\begin{enumerate}
\addtolength{\itemsep}{-0.7\baselineskip}
\item In the configuration $C_{\ref{updateCursor-check}}$ before the last execution of line \ref{updateCursor-check} inside a call to updateCur\-sor($c$),
the local variables {\it invDel} and {\it invIns} are equal to $\gc$.{\it invDel} and $\gc$.{\it invIns} respectively. \label{inv-lem}
\item  A successful delete($c$) advances $\gc'$ to the next item in $\gL$ for all $\gc'$ such that $\gc'.item=\gc.item$ just before the linearization point of delete($c$).\label{abs-update}
A successful insertBefore($c$, $val$) does not change $\gc$.
\item realNode({\it c.node).absVal} = $\gc$.{\it item.absVal} at all configurations except between the linearization point of a move and setting $c.node$ on line 
 \ref{moveLeft-set-cursor-copy}, \ref{moveLeft-set-cursor-remove}, \ref{moveLeft-set-cursor} or \ref{moveRight-set-cursor}.\label{realNode-lem}
\item If $c.node$ is set to $v$ on line \ref{moveLeft-set-cursor-copy}, \ref{moveLeft-set-cursor-remove}, \ref{moveLeft-set-cursor} or \ref{moveRight-set-cursor} inside a move called with $c$,
between the linearization point of the move and setting $c.node$ on one of those lines, realNode({\it v).absVal} = $\gc$.{\it item.absVal}. \label{move-lem}
\item The sequence of reachable Nodes (excluding {\it head} and {\it tail}) and the sequence of items in $\gL$ have the same values and abstract values. \label{seq-lem}
\end{enumerate}
\end{lemma}
\noindent {\it Proof sketch.}
Statement \ref{inv-lem}:
$invDel$ is true at $C_{\ref{updateCursor-check}}$ if and only if $c.node$ points to a Node $v$ that is marked and unreachable at some earlier execution of line \ref{updateCursor-check}.
It can be shown using induction hypothesis \ref{realNode-lem} that this is true if and only if $\gc$ was invalidated when $v$ was deleted by a delete($c'$) after the linearization point of the previous operation called with $c$
such that $c \ne c'$ and $\gc.item=\gc'.item$ at the linearization point of delete($c'$).
The proof for $invIns$ is similar.

Statement \ref{abs-update}:
Suppose $C$ is the configuration before a successful forward CAS of an Info object $I$ created by a delete($c$).
Let $C_{\ref{updateCursor-check}}$ be the configuration before the last execution of line \ref{updateCursor-check} inside the call to updateCursor on line \ref{del-call-updateCursor} preceding the creation of $I$.
By Lemma \ref{reach-lem}, realNode({\it c.node})={\it c.node} at $C_{\ref{updateCursor-check}}$. 
So, at $C_{\ref{updateCursor-check}}$, $\gc.item.absVal=c.node.absVal$ (by induction hypothesis \ref{realNode-lem}) and $\gc$.{\it invDel} is false (by induction hypothesis \ref{inv-lem}).
Since $c.node = I.nodes[1]$ is reachable at $C$ (by Lemma \ref{ptr-lem}.\ref{ptr-reach-lem}), 
$\gc .item.absVal$ is still $c.node.absVal$ at $C$ (by induction hypothesis \ref{realNode-lem}) and $\gc.invDel$ is still false at $C$ (because $c.node$ has not been removed).
Since $\gc.item = \gc'.item$ at $C$, the forward CAS of $I$ advances $\gc'.item$ to the next item in $\gL$.
The proof for insertBefore($c$, $val$) is similar.

Statement \ref{realNode-lem}:
We consider different cases that change $\gc$, $c.node$ or realNode($c.node$).

Case 1:
$\gc$ or realNode is changed.
Only a successful forward CAS of an Info object $I$ can change~$\gc$ or realNode.
Suppose a delete($c$) created $I$. 
Then, the CAS changes $x_I.nxt$ from $y_I$ to $z_I$.
Just after the CAS, $y_I$ is unreachable and marked, $y_I.nxt = z_I$ (by Lemma \ref{ptr-lem}.\ref{frozen-lem}) and $z_I$ is still reachable 
(step~2 of Fig.~\ref{del-fig}).
So, if realNode($c'.node)=y_I$ before the CAS, then realNode($c'.node)=z_I$ after the CAS. 
Thus, if the CAS advances $\gc'$ to the next item in $\gL$, it also changes realNode($c'.node$) to the next reachable Node.
By induction hypothesis \ref{seq-lem},  claim is preserved.
An insertion's forward CAS is similar.

Case 2: 
line \ref{ins-set-cursor}, \ref{del-set-cursor},  \ref{updateCursor-set-new-copy} or \ref{updateCursor-set-nxt} sets $c.node$.
When line \ref{del-set-cursor} or \ref{updateCursor-set-nxt} changes $c.node$ from $u$ to $v$,
$u$ is marked and unreachable and $u.nxt=v$, so realNode($c.node$) is not changed.
Line \ref{ins-set-cursor} or~\ref{updateCursor-set-new-copy} are similar.

Case 3:
line \ref{moveLeft-set-cursor-copy}, \ref{moveLeft-set-cursor-remove}, \ref{moveLeft-set-cursor} or \ref{moveRight-set-cursor} sets $c.node$.
By induction hypothesis \ref{move-lem}, the claim is true.

Statement \ref{move-lem}:
By Lemma \ref{mr-lin} and \ref{ml-lin}, $\gc.item.absVal = v.absVal$ at the linearization point of the move.
The argument that all other steps preserve this claim is similar to Case 1 of Statement \ref{realNode-lem}.

Statement \ref{seq-lem}:
By induction hypothesis \ref{inv-lem}, unsuccessful updates change neither $\gL$ nor the reachable Nodes.
By Lemma \ref{ptr-lem}.\ref{first-ptr}, $\gL$ and the reachable Nodes are changed only by the first forward CAS of an Info object.
Let $C$ and $C'$ be the configurations before and after the successful forward CAS of $I$ created by a delete($c$).
(A similar argument applies to insertBefore.)
Since $c.node = I.nodes[1]$ is reachable at $C$ (by Lemma \ref{ptr-lem}.\ref{ptr-reach-lem}), $\gc.item.absVal = I.nodes[1].absVal$ at $C$ (by induction hypothesis~\ref{realNode-lem}).
Only $I.nodes[1]$ becomes unreachable at $C'$ (see Fig.~\ref{del-fig}).
Likewise, only $\gc.item$ is removed from $\gL$ at~$C'$.
\qed

\mysection{Amortized Analysis}
A cursor is {\it active} 
if it has been created, but  not yet destroyed.
Let $\dot{c}(op)$  be the maximum number of active cursors at any configuration during operation $op$. 
We prove that the amortized complexity of each update $op$ is $O(\dot{c}(op))$ and each move is $O(1)$.
More precisely, for any finite execution $\alpha$, the total number of steps performed in $\alpha$ is $O(\sum_{\text{$op$ is an update in $\alpha$}}\dot c(op) + \text{number of move operations in $\alpha$})$.
It follows that the implementation is non-blocking.
The complete analysis (available in \cite{tech-rep}) is quite complex, so we only
sketch it here.
Parts of it are similar to the analysis of search trees by Ellen et al.~\cite{AA-BST} but the parts dealing with cursors and moves are original. 
They used a combination of an aggregate analysis and an accounting method argument. 
We simplified the analysis using the potential method and show how to generalize their argument to handle operations that flag more than two nodes.

Each iteration of the loop at line \ref{ins-start-loop}--\ref{ins-return-true} or \ref{del-start-loop}--\ref{del-return-true} inside an update is called an {\it attempt}. 
A complete attempt is {\it successful} if it returns on line \ref{ins-return-invalid}, \ref{ins-return-true}, \ref{del-return-invalid} or \ref{del-return-true}; otherwise it is {\it unsuccessful}.
Each iteration of the loop at line \ref{updateCursor-check}--\ref{updateCursor-end-loop} and each attempt (excluding the call to updateCursor) take $O(1)$ steps, so we assume they take one unit of time.
We design a potential function $\Phi$ that is the sum of three parts, $\Phi_{cursor}, \Phi_{state}$ and $\Phi_{flag}$ to satisfy the following properties.
\begin{compactitem}
\item
Each iteration of line \ref{updateCursor-check}--\ref{updateCursor-end-loop} in updateCursor decreases 
$\Phi_{cursor}$ and does not affect $\Phi$ otherwise.
\item
Each move (excluding its call to updateCursor) does not change $\Phi$.
\item
Each unsuccessful attempt of an update (excluding the call to updateCursor) decreases $\Phi$.
\item
The final, successful attempt of an update operation $op$ increases $\Phi$ by at most $O(\dot{c}(op))$.
\end{compactitem}
We sketch the main ideas here; Appendix \ref{potential-func} gives more details and \cite{tech-rep} gives the complete argument.

$\Phi_{cursor}$ is 
used to bound the amortized complexity of updateCursor.
Roughly speaking, $\Phi_{cursor}$ is the sum of the lengths of the paths that would have to be traced from each cursor $c$'s location by the next call to updateCursor($c$).
A successful forward CAS of an update $op$ adds  one to $\Phi_{cursor}$ for each cursor $c$ that will have to perform an additional
iteration of line \ref{updateCursor-check}--\ref{updateCursor-end-loop} in updateCursor($c$).
This adds at most $\dot{c}(op)$ to the amortized cost of $op$.
Since each iteration of updateCursor decreases $\Phi_{cursor}$ by 1, the amortized cost of updateCursor is 0.
Besides its call to updateCursor, a move only performs $O(1)$ steps, which we show do not 
affect $\Phi$.
It follows that a move's amortized complexity is~$O(1)$.

It remains to show
that the amortized number of failed attempts per update $op$ is $O(\dot{c}(op))$.

An attempt can fail if at line \ref{checkInfo-check-state}
it reads marked or copied from the state of
one of the nodes it wants to flag 
(indicating that the node is no longer in the list).
We use $\Phi_{state}$ to bound the number of attempts 
that fail in this way. 
When an update $op$ sets the state of a node, it adds $O(\dot{c}(op))$
units to $\Phi_{state}$ to pay for the attempts that may fail as a result.
CAS steps that change  list pointers also store potential in $\Phi_{state}$, 
because they may change the nodes that other updates wish to flag.

An attempt $att$ of an update may also fail because one of the nodes it wishes to flag gets flagged by an attempt $att'$ of another operation.
(Then, $att$'s test at line \ref{checkInfo-check-inProgress} or \ref{checkInfo-return-false3} fails or $att$ fails to flag a node on line \ref{help-flag}.)
If $att'$ were guaranteed to succeed in this case, the analysis would be simple.
However, $att'$ itself may also fail because it is blocked by the attempt of some third operation, and so on.
We employ $\Phi_{flag}$ to bound the amortized number of attempts that fail in this way by modifying the approach used for trees in \cite{AA-BST}.
The definition of $\Phi_{flag}$ is intricate, but it has the following properties:
\begin{compactitem}
\item The invocation of an update $op$ increases $\Phi_{flag}$ by $O(1)$ for each pending update (total of~$O(\dot{c}(op))$).
\item When the $status$ of an Info created by $op$ is set to committed, it increases $\Phi_{flag}$ by~$O(\dot c(op))$.
\item When the first flag CAS of an Info object on a node fails, it decreases $\Phi_{flag}$ by~1.
\item When the test at line \ref{checkInfo-check-inProgress} or \ref{checkInfo-return-false3} fails, it decreases $\Phi_{flag}$ by~1.
\end{compactitem}

It follows that the amortized cost of unsuccessful attempts is 0 and the amortized cost of the last attempt of update $op$ is~$O(\dot c(op))$.

\mysection{Conclusion}
The amortized bound of $O(\dot c(op))$ for an update $op$
is quite pessimistic: the worst case would happen only if many overlapping updates
are scheduled in a very particular way.
We expect our list would have even better performance in practice.
Preliminary experimental results suggest that our list scales well in a multicore system.
(See Appendix~\ref{exp-result}.)
In particular, it greatly outperforms an implementation using transactional memory,
which has more overhead than our handcrafted implementation.

Though moves have constant amortized time, they are not wait-free.
For example, if cursors $c$ and $c'$ point to the same node, a moveLeft$(c)$ may never terminate if an infinite sequence of insertBefore($c'$) operations succeed, because the updateCursor routine called by the move could run forever.

Future work includes thorough experimental evaluation and designing shared cursors. 
Generalizing our coordination scheme could provide a simpler way to design non-blocking data structures.
Although the proof of correctness and analysis is complex, it is modular, so it could be applied more generally.



\noindent {\bf \\Acknowledgments.}
I would like to thank my supervisor, Eric Ruppert, for his great guidance, advice and tremendous support and Michael L. Scott for giving us access to his multicore machines.


\bibliographystyle{plain}
\bibliography{ref}


\appendix
\mysection{Formal Sequential Specification}
\label{formal-spec}


A list of items $(L, S)$ is a pair that supports eight types of operations: createCursor, destroyCursor, resetCursor, insertBefore, delete,  get, moveRight and moveLeft.
$L$ is a finite sequence of distinct items ending with special end-of-list marker (EOL), and $S$ is a set of cursors.
Each cursor is a tuple $(name, item, invDel, invIns, id)$ that includes the name of the cursor, the item that the cursor is associated with, 
two boolean values and the id of the process that owns the cursor.
The item and the bits of a cursor $c$ are denoted $c.item$, $c.invDel$ and $c.invIns$ respectively. 
A process $p$ can call an operation with a cursor $c$ only if $p$ itself created $c$ and $c$ has not been destroyed.

The state of the list is initially $(\langle$EOL$\rangle, \emptyset)$.
We describe the state transitions and responses for each type of operation on a list in state $(L, S)$.
Let $firstItem$ be the first item in $L$.
If $c$ is a cursor in $S$, let $c.nxtItem$ be the next item after $c.item$ in $L$ (if it exists) and $c.prvItem$ be the item preceding $c.item$ in $L$ (if it exists).
If an operation is called with a cursor $c$, the operation sets both $c.invDel$ and $c.invIns$ to false before the operation terminates.

- createCursor() called by process $p$ adds the tuple $(name$, $firstItem$, false, false, $p)$ to the set $S$ and returns ack.

- destroyCursor($c$) removes $c$ from $S$ and returns ack.

- resetCursor($c$) sets $c.item$ to $firstItem$ and returns ack.

If delete($c$), get($c$), moveRight($c$) or moveLeft($c$) is called and $c.invDel$ is true, the operation returns invalidCursor. 
If insertBefore($c$) is called and either $c.invDel$ or $c.invIns$ is true, the operation returns invalidCursor. 

Otherwise, the operation induces the following state transition and response.

- insertBefore($c$, $val$) adds a new item with value $val$ just before $c.item$ in $L$ and returns true.
For all cursors $c' \ne c$ such that $c'.item = c.item$, it sets $c'.invIns$ to true.

- delete($c$), if $c.item \ne$ EOL, removes $c.item$ from $L$. 
For all cursors $c' \ne c$ such that $c'.item = c.item$, it sets $c'.item$ to $c.nxtItem$ and $c'.invDel$ to true. 
It also sets $c.item$ to $c.nxtItem$ and returns true.

If $c.item =$ EOL, delete($c$) returns false. 


- get($c$) does not change $(L, S)$ and it returns the value of $c.item$.

- moveRight($c$) does not change $L$.
If $c.item \ne$ EOL, it sets $c.item$ to $c.nxtItem$ and returns true;
otherwise, it does not change $(L, S)$ and returns false.

- moveLeft($c$) does not change $L$.
If $c.item \ne firstItem$, it sets $c.item$ to $c.prvItem$ and returns true;
otherwise, it does not change $(L, S)$ and returns false.

\mysection{Potential Function Used in Amortized Analysis}
\label{potential-func}
Here, we define the potential function  that is used in our amortized analysis.
Let $(L,S)$ be a pair that represents the list, $L$ is a sequence of items and $S$ is a set of cursors.
Let $c$ be a cursor in $S$, $u$, $v$ and $w$ be nodes. 
Our potential function $\Phi$ consists of three parts $\Phi_{cursor}$, $\Phi_{flag}$ and $\Phi_{state}$, which we define in~turn.

First, we define $\Phi_{cursor}$.
Intuitively, potential is stored in $\Phi_{cursor}$ by the successful forward CAS of an update to pay for the resulting updates to other cursors during updateCursor later.


\begin{eqnarray*}
realNode(u) & = & \left\{
 \begin{array}{l l l}
realNode(u.copy) & \quad \mbox{if $u.state =$ copied and $u$ is unreachable},\\
realNode(u.nxt) & \quad \mbox{if $u.state =$ marked and $u$ is unreachable},\\
u & \quad \text{otherwise}.
  \end{array}
  \right.\\
length(u)& =& \left\{
 \begin{array}{l l l}
length(u.copy)+1 & \quad \text{if $u.state =$ copied and $u$ is unreachable},\\
length(u.nxt)+1 & \quad \text{if $u.state =$ marked and $u$ is unreachable},\\
0 & \quad \text{otherwise}.
  \end{array}
  \right. \\
\phi_{cursor}(c) &=& \left\{
 \begin{array}{l l l}
length(u) & \quad \text{between the linearization point of a move called with} \\
& \quad \text{$c$ and setting $c.node$ to $u$ on line \ref{moveLeft-set-cursor-copy}, \ref{moveLeft-set-cursor-remove}, \ref{moveLeft-set-cursor} or \ref{moveRight-set-cursor}}\\
& \quad \text{if the move sets $c.node$ on line \ref{moveLeft-set-cursor-copy}, \ref{moveLeft-set-cursor-remove}, \ref{moveLeft-set-cursor} or \ref{moveRight-set-cursor}},\\
length(c.node) & \quad \text{otherwise}.
 \end{array}
  \right. \\
\Phi_{cursor} & = & \sum_{c \in S} \phi_{cursor}(c).  \\ 
\end{eqnarray*}

Next, we define the function $\Phi_{flag}$.
Intuitively, potential is stored in $\Phi_{flag}$ by successful flag CAS steps to pay for unsuccessful flag CAS steps and attempts whose calls to checkInfo later return on line \ref{checkInfo-return-false1} or \ref{checkInfo-return-false3}.
In addition, potential is stored in $\Phi_{flag}$ by setting the $status$ of Info objects to committed to pay for successful flag CAS steps.
Let $op$ be an active update operation that is called with $c$.\\ 

\indent $node_1(op)$ = $realNode$($c.node$) when $c$ is the cursor with which $op$ was invoked\\ 
\indent $node_0(op)$ = the reachable node whose $nxt$ pointer is $node_1(op$) \\
\indent $node_2(op)$ = the node that $node_1(op).nxt$ points to \\

For $i=0$, 1, 2, $lose_i(op)$ is initially set to 3 when $op$ is invoked and is updated as follows.

\[
 \begin{array}{l l l}
\text{set to 3} & \quad \text{when a forward or backward CAS succeeds}\\
\text{set to 3} & \quad \text{when some other operation sets the $info$ of $node_i(op$)} \\
\text{set to 2} & \quad \text{when the first flag CAS of $I$ created by $op$ on $I.nodes[i]$ fails}\\
\text{decremented} & \quad \text{when $lose_i(op) > 0$ and the read 
of $oldInfo[i].status$ in $op$'s line \ref{checkInfo-check-inProgress}} \\ & \quad \text{reads inProgress}\\
\text{decremented} & \quad \text{when $lose_i(op) > 0$ and the read of $nodes[i].info$ in $op$'s line \ref{checkInfo-return-false3} reads} \\ & \quad \text{a value different from $oldInfo[i]$.}\\
&\\
  \end{array}
\vspace{-10pt}
   \]



Let 
\[\mathrm{flag}(u) = \left\{
 \begin{array}{l l l}
1 & \quad \text{if $u.info.status$ is inProgress},\\
0 & \quad \text{otherwise}.
  \end{array}
\vspace{-10pt}
  \right. \]

We define an auxiliary variable abort($u$) that is initially 0 and updated as follows.
\[
 \begin{array}{l l l}
\text{set to 1} & \quad \text{when a flag CAS on $u$'s successor succeeds.}\\
\text{set to 1} & \quad \text{when a forward CAS changes $u.nxt$.}\\
\text{set to 0} & \quad \text{when $u.info.status$ is changed from inProgress to committed or aborted.}\\
&\\
  \end{array}
\vspace{-10pt}
 \]

Let  $\dot u$ at a configuration be the number of  updates running at that configuration.\\

\begin{eqnarray*}
\phi_{flag}(v) &=& \sum_{\text{$w$ is after $v$ in the list, including $v$}} (\mbox{abort}(w) - \mbox{flag}(w))\\
\Phi_{flag}& = &\sum_{op} (3 \cdot \sum\limits_{i=0}^2 \phi_{flag} (node_i(op)) + \sum\limits_{i=0}^2 lose_i(op)) + 27 \cdot \dot{u}^2
\end{eqnarray*}
where the sum is taken over all active update operations $op$.\\


By definition, $lose_i(op$) is never negative.
Moreover, at any one time, at most 3 nodes might be flagged by an Info object created by $op$ and this could contribute $-3 \dot u$ to each  $\phi_{flag}(v)$ and hence $-27 \dot u$ to $3 \cdot \sum\limits_{i=0}^2 \phi_{flag} (node_i(op))$ and 
$-27 \dot u^2$ to $\Phi_{flag}$.
The addition of the term  $27 \dot u^2$ ensures that 
$\Phi_{flag}$ is never negative.
The invocation of an update increases $\dot u$ by 1. 
 So, the invocation of an update increases $\Phi_{flag}$ by at most $27(\dot u^2 - (\dot u - 1)^2) + 9 = 54 \cdot \dot u - 18$.  
Since each update is called with a distinct cursor, $\dot u \le \dot c(op)$.
Thus, the invocation of an update contributes $O(\dot c(op))$ to $\Phi_{flag}$.


Next, we define the function $\Phi_{state}$.
Intuitively, potential is stored in $\Phi_{state}$ by successful forward and backward CAS steps, setting the $state$ of Node objects and update operations' invocations to pay for attempts that return on line \ref{checkInfo-return-false2}  later.
The $\phi_{state}(op)$ is initially 2 when $op$ is invoked and is updated as follows.

\[
 \begin{array}{l l l}
\text{set to 2} & \quad \text{when a forward or backward CAS succeeds}\\
\text{set to 2} & \quad \text{when the $state$ of some node is changed from ordinary to marked or copied}\\  
\text{decremented} & \quad \text{when $\phi_{state}(op) > 0$ and $op$ reads marked or copied from a node's $state$ field}\\ & \quad \text{on line \ref{checkInfo-check-state}.}\\
&\\
\end{array}
\vspace{-10pt}
\]

\begin{eqnarray*}
\Phi_{state} = \sum\limits_{op} \phi_{state}(op)
\end{eqnarray*}

where the sum is taken over all active update operations $op$.\\

For our analysis, we use the sum of the three potential functions we have defined.  \\

\begin{eqnarray*}
\Phi = \Phi_{cursor} + \Phi_{flag} + \Phi_{state}
\end{eqnarray*}

If an operation takes a step that is not inside the help routine, we say the step {\it belongs} to the operation.
Let $I$ be an Info object created by update operation $op$.
We say that any step inside any call to help($I$) {\it belongs} to~$op$.
The following tables show the change in the potential functions caused by the steps that belong 
to an operation. 
The detailed proofs of these claims are available in \cite{tech-rep}.
First, we have the changes to the potential function within updateCursor.
In the following table, $\Delta \Phi_{x}$ shows the changes to $\Phi_{x}$ by a call to updateCursor.

\begin{table}[H]
\begin{center}
\begin{tabular}{|l|l|l|l|}
\hline
\rowcolor{lightgray}
step & $\Delta \Phi_{cursor}$ & $\Delta \Phi_{flag}$ & $\Delta \Phi_{state}$ \\ \hline
line \ref{updateCursor-set-new-copy} and \ref{updateCursor-set-nxt}&-1&0&0 \\ \hline
\end{tabular}
\end{center}
\caption{updateCursor}
\end{table}

A complete attempt $att$ fails if $att$'s call to checkInfo on line \ref{ins-call-checkInfo} or \ref{del-call-checkInfo} or $att$'s call to help on line \ref{ins-call-help} or \ref{del-call-help} returns false.
For an Info object $I$, if help($I$) returns false, the first flag CAS of $I$ on $I.nodes[i]$ (for some $i$) fails.
The next six tables show the changes to $\Phi$ by unsuccessful attempts of updates.
In the remaining tables, $\Delta \Phi_{x}$ shows the changes to $\Phi_{x}$ due to steps belonging to the attempt, excluding its call to updateCursor..

\begin{table}[H]
\begin{center}
\begin{tabular}{|l|l|l|l|}
\hline
\rowcolor{lightgray}
step & $\Delta \Phi_{cursor}$ & $\Delta \Phi_{flag}$ & $\Delta \Phi_{state}$ \\ \hline
line \ref{checkInfo-check-inProgress} reads inProgress&0&$-1$&0 \\ \hline
\end{tabular}
\end{center}
\caption{checkInfo returns false on line \ref{checkInfo-return-false1}}
\end{table}

\begin{table}[H]
\begin{center}
\begin{tabular}{|l|l|l|l|}
\hline
\rowcolor{lightgray}
step & $\Delta \Phi_{cursor}$ & $\Delta \Phi_{flag}$ & $\Delta \Phi_{state}$ \\ \hline
line \ref{checkInfo-check-state} reads copied or marked&0&0&$-1$ \\ \hline
\end{tabular}
\end{center}
\caption{checkInfo returns false on line \ref{checkInfo-return-false2}}
\end{table}

\begin{table}[H]
\begin{center}
\begin{tabular}{|l|l|l|l|}
\hline
\rowcolor{lightgray}
step & $\Delta \Phi_{cursor}$ & $\Delta \Phi_{flag}$ & $\Delta \Phi_{state}$ \\ \hline
line \ref{checkInfo-return-false3} reads the $info$ field&0&$-1$&0 \\ \hline
\end{tabular}
\end{center}
\caption{checkInfo returns false on line \ref{checkInfo-return-false3}}
\end{table}

\begin{table}[H] 
\begin{center}
\begin{tabular}{|l|l|l|l|}
\hline
\rowcolor{lightgray}
step & $\Delta \Phi_{cursor}$ & $\Delta \Phi_{flag}$ & $\Delta \Phi_{state}$ \\ \hline
line \ref{help-flag} fails to flag the first node&0&$-1$&0 \\ \hline
line \ref{help-set-aborted} sets the $status$ to aborted&0&0&0 \\ \hline
\end{tabular}
\end{center}
\caption{attempt fails when it fails to flag the first node}
\label{table-flag-1}
\end{table}

\begin{table}[H]
\label{table-flag-2}
\begin{center}
\begin{tabular}{|l|l|l|l|}
\hline
\rowcolor{lightgray}
step & $\Delta \Phi_{cursor}$ & $\Delta \Phi_{flag}$ & $\Delta \Phi_{state}$ \\ \hline
line \ref{help-flag} flags the first node&0&$\le -3$&0 \\ \hline
line \ref{help-flag} fails to flag the second node&0&$-1$&0 \\ \hline
line \ref{help-set-aborted} sets the $status$ to aborted&0&0&0 \\ \hline
\end{tabular}
\end{center}
\caption{attempt fails when it fails to flag the second node}
\label{table-flag-2}
\end{table}

\begin{table}[H]
\label{table-flag-3}
\begin{center}
\begin{tabular}{|l|l|l|l|}
\hline
\rowcolor{lightgray}
step & $\Delta \Phi_{cursor}$ & $\Delta \Phi_{flag}$ & $\Delta \Phi_{state}$ \\ \hline
line \ref{help-flag} flags the first node&0&$\le -3$&0 \\ \hline
line \ref{help-flag} flags the second node&0&$\le -3$&0 \\ \hline
line \ref{help-flag} fails to flag the third node&0&$-1$&0 \\ \hline
line \ref{help-set-aborted} sets the $status$ to aborted&0&0&0 \\ \hline
\end{tabular}
\end{center}
\caption{attempt fails when it fails to flag the third node}
\label{table-flag-3}
\end{table}

The following table shows the changes to $\Phi$ when a call to the help routine returns true.

\begin{table}[H]
\begin{center}
\begin{tabular}{|p{7cm}|l|l|l|}
\hline
\rowcolor{lightgray}
step & $\Delta \Phi_{cursor}$ & $\Delta \Phi_{flag}$ & $\Delta \Phi_{state}$ \\ \hline
line \ref{help-flag} flags the first node&0&$\le -3$&0 \\ \hline
line \ref{help-flag} flags the second node&0&$\le -3$&0 \\ \hline
line \ref{help-flag} flags the third node&0&$\le -3$&0 \\ \hline
line \ref{help-set-marked} or \ref{help-set-copied} changes the $state$ of a node from ordinary to marked or copied&0&0&$\le 2 \cdot \dot{c}(op)$ \\ \hline
line \ref{help-forward-CAS} succeeds&$\le \dot{c}(op)$& $\le 63 \cdot \dot{c}(op)$&$\le 2 \cdot \dot{c}(op)$ \\ \hline
line \ref{help-backward-CAS} succeeds&0&$\le 9 \cdot \dot{c}(op)$&$\le 2 \cdot \dot{c}(op)$ \\ \hline
line \ref{help-set-committed} changes $status$ of the Info from ordinary to committed&0&$\le 27 \cdot \dot{c}(op)$&0 \\ \hline
line \ref{ins-set-cursor} or \ref{del-set-cursor}&$-1$&0&0 \\ \hline

\end{tabular}
\end{center}
\caption{the call to help on line \ref{ins-call-help} or \ref{del-call-help} returns true}
\end{table}

\mysection{Preliminary Empirical Results}
\label{exp-result}





Here, we have preliminary evaluation of our implementation on a multicore system to show our implementation is scalable and practical.
We evaluated our implementation (NBDLL) on a Sun SPARC Enterprise T5240 with 32GB RAM and two UltraSPARC T2+ processors, each with eight 1.2GHz cores, for a total of 128 hardware threads. 
The experiments were run  in Java. 
The Sun JVM version 1.7.0\_3 was run in server mode.
The heap size was set to 4G to ensure that the garbage collector was invoked regularly, but not too often.
We focus on testing the scalability of our list.
We also compare NBDLL to a doubly-linked list using the Java implementation of transactional memory of \cite{JavaSTM} (STMDLL). 
In each graph, the x-axis is the number of threads (from 1 to 128) and
each data point 
is the average of fifteen 4-second trials.
Error bars show  standard deviations.
Since  Java  optimizes running code, 
we ran two warm-up trials before each experiment.

\begin{figure}[h]
\begin{minipage}{0.53 \textwidth}
\centering
\includegraphics[width=\textwidth]{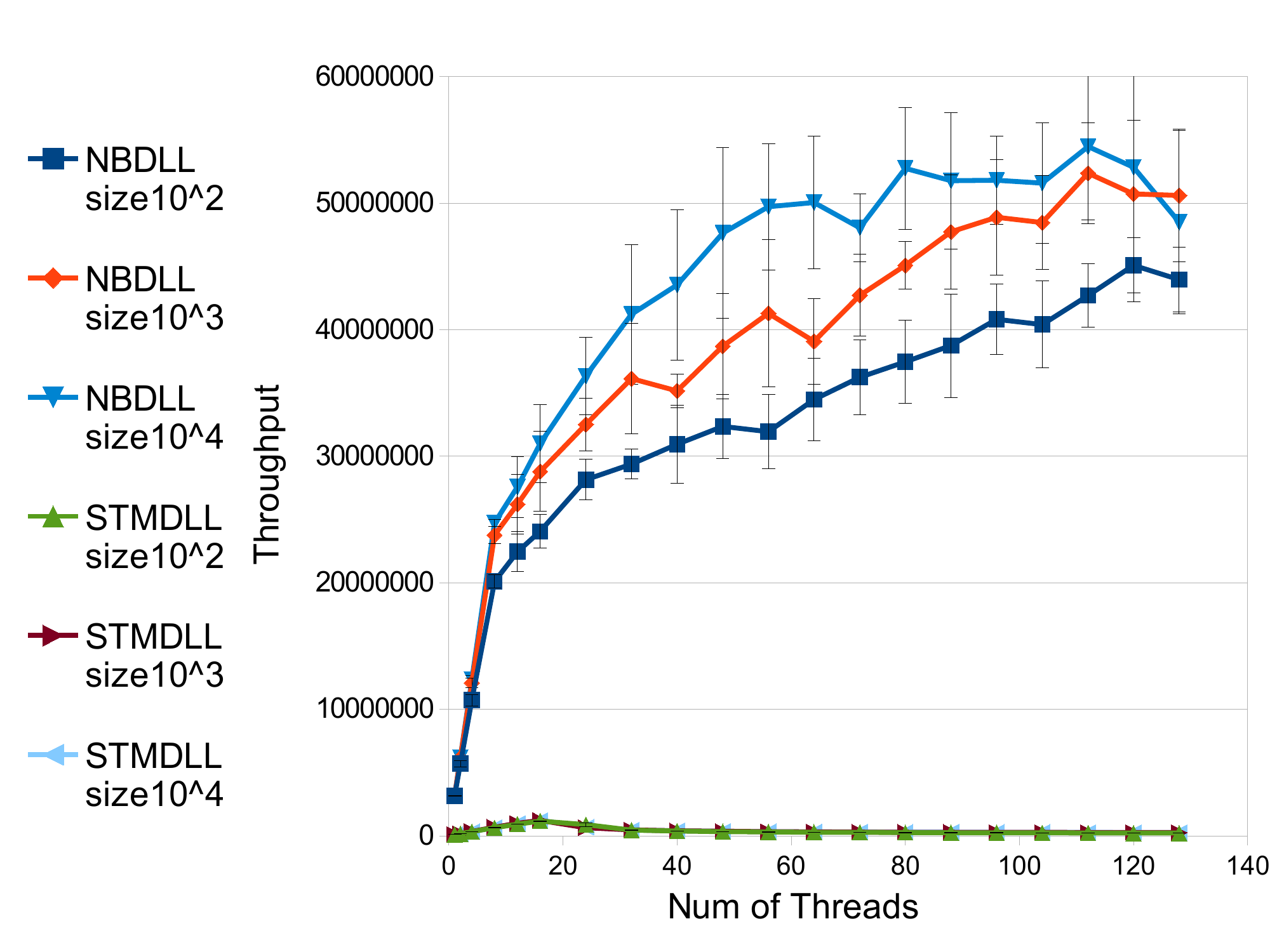}
\vspace{-30pt}
\caption{ratio: i5-d5-m90}
\label{chart-5-5-90}
\end{minipage}
\begin{minipage}{0.47 \textwidth}
\centering
\includegraphics[width=\textwidth]{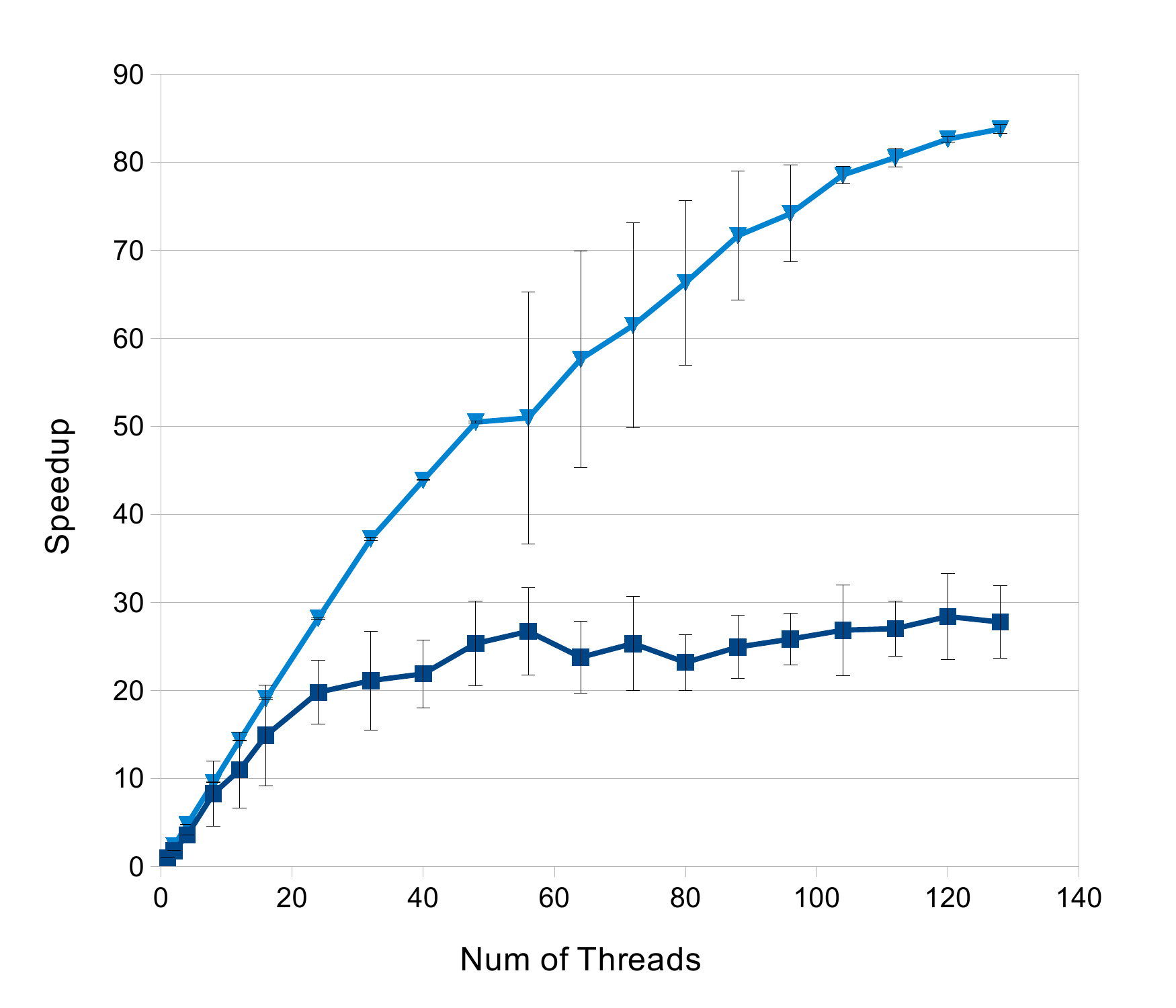}
\vspace{-30pt}
\caption{sorted list}
\label{chart-sorted}
\end{minipage}
\vspace{0pt}
\end{figure}

In the first scenario, 
we ran NBDLL and STMDLL with operation ratios of 
 5\% insertBefores, 5\% deletes and 90\% moves (i5-d5-m90). 
We ran the experiments with three different list sizes:
$10^2$, $10^3$ and $10^4$ to measure performance under high, medium and low contention.
(Other ratios are tested gave similar results.)
(See Fig.~\ref{chart-5-5-90}.)
The y-axis in Fig.~\ref{chart-5-5-90} gives throughput (operations per second). 
Each process's cursor had a random starting location.  
To increase the contention consistently when the number of threads are increased, we try to keep the size of the list and the distribution of the cursors consistent through the experiments.
We chose fractions of moveLefts and moveRights so that the cursors remained approximately evenly distributed across the list. 
Each process alternated between insert and delete to keep the list length roughly constant.
Our results show that NBDLL scales much better than STMDLL.
NBDLL scales best for up to 16 threads (since the machine has 16 cores).
For the list with $10^2$ elements, throughput scales more slowly since contention becomes very~high.

In the second scenario, we implemented a sorted list. 
(See Fig.~\ref{chart-sorted}.)
In Fig.~\ref{chart-sorted}, speedup is the throughput of key insertions and deletions (which consists of many move operations and zero or one update) over the throughput of one process.
Threads insert or delete random keys from the ranges $[0,2\cdot10^2]$ and $[0,2\cdot10^4]$ and the list is initialized to be half-full.
Since the number of move operations called to find the location for insertion and deletion depends on the size of the list, 
it is not fair to compare the throughput of lists with different sizes.  
Since speedup compares the number of updates performed by all threads to one thread, we have speedup of lists with different sizes in Fig.~\ref{chart-sorted} instead of throughput. 
For shorter lists, less time is required to find the correct location, but contention is high. 
As our results show, our implementation scales well and longer lists scale better because of lower contention.


\end{document}